\documentclass{elsart}

\usepackage{amssymb}
\usepackage{amsmath}
\usepackage{graphicx}

\begin{document}

\begin{frontmatter}



\title{Direct string magnetic gradiometer for space applications}


\author[uwa]{Andrew Sunderland\corauthref{cor1}}\ead{asund@physics.uwa.edu.au}, \author[uwa,gravitec]{Alexey V Veryaskin}, \author[uwa,gravitec]{Wayne McRae}, \author[uwa]{Li Ju}, \author[uwa]{David G Blair}
\address[uwa]{School of Physics, University of Western Australia, Perth, WA, Australia}
\address[gravitec]{Gravitec Instruments, Perth, WA, Australia}
\corauth[cor1]{Corresponding author. Tel.: +61 422 282 438.}

\begin{abstract}
Recently, a novel Direct String Magnetic Gradiometer (DSMG) has been
developed, where a vibrating wire, driven by an AC current, is used
as a single sensitive element. It is designed to directly measure
the local off-diagonal components of the magnetic gradient tensor,
Bxz, Byz and Bxy, provided the distance to an object creating
magnetic anomalies is much larger than the length of the string.
This requirement is well satisfied in space, if the sensor is
deployed from a satellite platform orbiting near the planet under
surveillance. Current instruments operating at $1\mbox{ kPa}$
pressure achieve sensitivity of $1.8 \times 10^{-10}\mbox{
T}/\mbox{m}$ in the band $0.0001\mbox{ Hz}$ to $0.1\mbox{ Hz}$. In
this paper we show that proposed modifications to the current
gradiometer design, specifically aimed at the deployment in space,
could have a magnetic gradient sensitivity better than
$10^{-13}\mbox{T}/\mbox{m}/\sqrt{\mbox{Hz}}$ in the frequency range
of interest for specific missions both for fundamental research and
for such applications as geophysical exploration on Mars and other
solar system planets. Also, by combining a few single-axis magnetic
gradiometer modules, it is possible to deploy a full tensor magnetic
gradiometer.
\end{abstract}

\begin{keyword}
magnetic gradients \sep gradiometry \sep magnetometry \sep current
carrying string
\end{keyword}
\end{frontmatter}


\section{Introduction}

Magnetic gradiometry, as a powerful tool for magnetic anomaly
mapping, have been discussed in literature in conjunction with
future planetary and deep space missions (see for example Alves and
Madeira\cite{alves}\cite{acuna}). Although conventional
magnetometers are more commonly deployed on satellites, interest is
growing in the use of magnetic gradiometers to extract data that
cannot be obtained from magnetic field measurements alone. Hastings
et al\cite{hastings} described several advantages of using a
magnetic gradiometer to directly measure magnetic gradients in
space. In their paper, some cryogenically cooled SQUID-based
magnetic gradiometer designs have been considered. SQUID-based
magnetic gradiometers are currently under development mainly for
airborne geophysical reconnaissance purposes
\cite{stolz}\cite{tilbrook}. On their own, SQUID gradiometers
provide a very high sensitivity to magnetic gradients in the
laboratory environment\cite{fagaly}, and require some additional
auxiliary equipment and a compensation technique when deployed from
a moving platform\cite{stolz}. Due to logistical difficulties, the
use of SQUIDs in space has been limited until
recently\cite{acuna}\cite{klinger}.

Many fluxgate magnetometers have been previously used in space with
sensitivities ranging from $10^{-12}T/\sqrt{\mbox{ Hz}}$ to
$10^{-11}T/\sqrt{\mbox{ Hz}}$\cite{primdahl}. They do not require
any cryogenic environment, and fall into a medium range on the
sensitivity scale compared to the SQUID-based magnetometers. In the
past few years, fluxgate gradiometers have been proposed for space
missions. The best performance reported to date is $9.3 \times
10^{-11}\mbox{ T}/\mbox{m}$ in the band $0.01\mbox{ Hz}$ to
$10\mbox{ Hz}$\cite{merayo}.

\section{String Magnetic Gradiometer}

Recently, a novel Direct String Magnetic Gradiometer (DSMG) has been
developed\cite{veryaskin}\cite{mcrae}\cite{golden}. It consists of a
single aluminium 6061 alloy string or wire (i.e. an object with
transverse dimensions much smaller than its longitudinal dimension).
The string is held under tension with its second harmonic
oscillation mode at $f_{0} \approx 850\mbox{ Hz}$. An AC current
tuned to the second harmonic is used to drive the string. This sets
the string into a resonant motion due to the Ampere force per unit
length:

\begin{equation}
\frac{\partial\mathbf{f}}{\partial z} = i_{s}\left[\mathbf{e}_{z}
\times \mathbf{B}\right]\mbox{sin}(\omega t)
\end{equation}

where $\mathbf{e}_{z}$ is a unit vector along the Z direction chosen
to point along the string's length, $i_{s}$ is the amplitude of the
AC drive current, $\omega$ is the string's drive angular frequency
and $\mathbf{B}$ is the magnetic induction vector.

When the string is a stretched thin flat ribbon\cite{golden}, the
resonant motion is strictly one-dimensional with its sensitivity
axis pointing perpendicular to the plane of motion. In this case,
the ribbon represents a one-dimensional mechanical harmonic
oscillator having an infinite number of resonant
modes\cite{veryaskin}:

\begin{align}
&\frac{d}{dt^{2}}X_{n}\frac{2}{\tau}\frac{d}{dt}X_{n} +
\omega_{n}^{2}X_{n} = \notag\\&\ \ \ \left[\frac{2}{\pi n}\left(1 -
(-1)^{n}\right)B_{y} - (-1)^{n}\frac{2l}{\pi
n}\frac{dB_{y}}{dz}\right]\frac{i_{s}}{\eta}\mbox{sin}(\omega t) +
N_{n}(t)\label{alexeyeqn}
\end{align}

It is assumed that the ribbon vibrates in the XOZ plane of its local
reference frame the origin of which is coincident with a lower clamp
point. The upper clamp point determines the ribbon's length $l$.
$X_{n}(t)$ is the amplitude of a n-mode mechanical displacement of
the ribbon from its unperturbed position aligned with Z axis. It is
also assumed that all non-linear terms can be ignored as, in fact,
the maximum possible mechanical displacements do not exceed the
nanometer scale\cite{anand}. In Eq. \ref{alexeyeqn}, $\eta$ is the
ribbon's mass per unit length, and $\tau$ is its mechanical
relaxation time, which is the same for all resonant modes of the
ribbon. $N_{n}(t)$ represents the fundamental thermal noise source
(in terms of acceleration noise) which sets an absolute limit on the
sensitivity of DSMGs. It has the following correlation function in
the white noise area\cite{veryaskin}:

\begin{equation}
\left<N_{n}(t_{1})N_{m}(t_{1})\right> = \frac{8kT}{\eta
l\tau}\delta_{nm}\delta(t_{1} - t_{2})
\end{equation}

where $k = 1.4 \times 10^{-23}\mbox{ J$/$K}$ is Boltzmann's constant
and $T$ is absolute temperature.

As it follows from Eq. \ref{alexeyeqn}, the magnetic gradient term
of the driving force is coupled only to even resonant modes, while
the conventional magnetic field term is coupled to the odd ones.

During operation the string oscillator is an integral part of a
dynamic feedback loop\cite{veryaskin2}. The dynamic properties of
such a complex system are different from those of a stand-alone
mechanical oscillator described in previous work\cite{veryaskin}. In
particular, there are a number of additional parameters that can
play a crucial role in creating an optimised DSMG with a possibility
to implement effective noise suppressing algorithms, such as
electronic cooling\cite{ritter} and pulse feedback modulation
technique\cite{veryaskin3}.

By its very nature, DSMG is a modulation-demodulation device. Like
fluxgate magnetic gradiometers, it is capable of detecting the quasi
DC magnetic gradients both in relative and in absolute units. The
detection method provides strong immunity to the uniform magnetic
field. Firstly the AC drive current does not couple strongly to the
uniform field. The second harmonic drive frequency naturally couples
to the magnetic gradient but is well off the drive frequency that
couples to the uniform magnetic field. Secondly, the mechanical
displacement detection is designed to preferentially detect the
second harmonic oscillations and suppress the fundamental mode
oscillations. A common mode rejection ratio of the order of $10^{7}$
is naturally achieved without any balancing technique. The
mechanical Q factor of the ribbon provides first stage amplification
of the signal to the level where an instrumental read-out noise is
lower than the fundamental thermal noise of the ribbon. The latter
determines the fundamental rms noise floor of a
DSMG\cite{veryaskin2}:

\begin{equation}
\frac{dB_{y}}{dz} = \frac{2\pi}{i_{s}}\sqrt{\frac{2\eta
kT^{*}}{l^{3}\tau t}}\label{sensitivity}
\end{equation}

where, for the current DSMG design, $l = 0.25\mbox{ m}$ is the
length of the ribbon, $\eta = 8 \times 10^{-6}\mbox{ kg$/$m}$ is the
mass per unit length, $\tau \approx 0.1\mbox{ s}$ is the relaxation
time at a pressure of $1\mbox{ kPa}$, $t = 5\mbox{ s}$ is the
measurement time and $T^{*}$ is the effective noise temperature. A
complete theory of operation of DSMGs is presented in another
paper\cite{veryaskin2}.

A DSMG designed to date, operates typically at $T = 300\mbox{ K}$ in
a $1\mbox{ kPa}$ vacuum. An inductive read-out system has been
developed in order to detect the gradient driven displacements of
the ribbon at a level of $6 \times
10^{-13}\mbox{m}/\sqrt{\mbox{Hz}}$. The achieved sensitivity is $1.8
\times 10^{-10}\mbox{ T$/$m}$ in an unshielded environment in the
band $0.0001\mbox{ Hz}$ to $0.1\mbox{ Hz}$.

Below we consider some possible ways of greatly reducing the DSMG's
thermal noise limiting factor by using some advantages of the
natural space-borne environment. We show that a DSMG specifically
designed for space-borne applications can be as sensitive as
SQUID-based devices without the requirement of using cryogenic
equipment. Also, by their nature, DSMGs should exhibit a very low
$1/f$ noise, allowing measurements within the $0.0002\mbox{ Hz}$ to
$0.17\mbox{ Hz}$ band specified by Hastings et al\cite{hastings}.

\section{Thermal noise and the mechanical quality factor}

During deployment in space, the DSMG would operate at pressures
between $10^{-4}\mbox{ Pa}$ at an altitude of $230\mbox{
km}$\cite{nasa2} and $10^{-8}\mbox{ Pa}$ on the moon\cite{landis}.
This means that despite the large surface area to mass ratio of the
ribbon, gas friction damping is negligible. Measurements of the
intrinsic mechanical $Q$ factor of Aluminium 6061-T6511 by
Duffy\cite{duffy} with a cylinder of diameter 6mm give $Q$ factors
as high as $2.8 \times 10^{5}$ at $300\mbox{ K}$. Increasing the $Q$
factor of the ribbon is one way to reduce thermal noise.

The aluminium ribbon is strained and clamped at both ends. Initially
the amount of stress in tension is proportional to the amount of
strain. The tension sets the resonant frequency of the magnetic
gradiometer. Low levels of stress relaxation are desirable to ensure
a long operating life for the DSMG. Table \ref{stress} shows that
annealed aluminium is not a viable material for the ribbon due to
excessive levels of stress relaxation.

For space deployment it is proposed to use a wide thin ribbon. This
would maximise the surface area available for thermal radiation
which would allow a higher current to be pumped along the ribbon and
yet minimise the mass per unit length. The high current and low mass
would increase the sensitivity as per Eq. \ref{sensitivity}.

The mechanical $Q$ factor of a very thin ribbon is lower than the
$Q$ factor of the bulk metal because of surface losses. Gretarsson
et al give an expression relating $Q_{ribbon}$ to
$Q_{bulk}$\cite{gretarsson}:

\begin{equation}
\frac{1}{Q_{ribbon}} = \frac{1}{Q_{bulk}}\left(1 +
\mu\frac{d_{s}}{V/S}\right)\label{surface}
\end{equation}

where $V/S$ is the volume to surface ratio, $d_{s}$ is the
dissipation depth of surface loss and $\mu$ is a measure of the
fraction of elastic energy attributable to strains at the surface of
the sample. Preliminary experiments by the authors on aluminum 6061
with a ribbon of width $0.125\mbox{ mm}$ and thickness $0.025\mbox{
mm}$ in a vacuum of $10^{-2}\mbox{ Pa}$ have measured a ribbon Q
factor of $200 \pm 20$. In such a thin ribbon the surface loss is
dominant. Fitting $Q_{ribbon} = 200$ to Eq. \ref{surface} gives:

\begin{equation}
\frac{1}{Q_{ribbon}} = \frac{1}{Q_{bulk}}\frac{2\mu d_{s}}{t}
\end{equation}

where $Q_{bulk} = 280000$ is Duffy's result\cite{duffy}, $\mu = 1$,
$d_{s} = 16\mbox{ mm}$ and $t = 0.025\mbox{ mm}$ is the thickness of
the ribbon.

When tension is applied to the ribbon, a portion of the vibration
energy is stored as tensile stress instead of surface strains. This
can lead to a higher $Q$ factor, also known as an enhanced $Q$
factor. Using a formula in Gonzalez et al\cite{gonz}, the value of
$\mu$ for the second violin mode is:

\begin{equation}
\mu = \frac{2}{l}\sqrt{\frac{IE}{T}}\label{mu}
\end{equation}

where $l$ is the length of the ribbon, $I = \frac{t^{3}w}{12}$ is
the moment of area, $w$ is the width of the ribbon, $E$ is the
young's modulus of the aluminum alloy and $T$ is the tension.
Substituting in the angular resonant frequency of the second violin
mode $\omega = \frac{n\pi}{l}\sqrt{\frac{T}{\rho tw}}$ into Eq.
\ref{mu} gives:

\begin{equation}
\mu = \frac{4n\pi t}{\omega l^{2}}\sqrt{\frac{3E}{\rho}}\label{mu2}
\end{equation}

where $\rho = 2690\mbox{ kgm$^{-3}$}$ is the density of aluminum
6061 and $n = 2$ is the mode number. The ribbon $Q$ factor is then:

\begin{equation}
\frac{1}{Q_{ribbon}} = \frac{1}{Q_{bulk}}\frac{8n\pi d_{s}}{\omega
l^{2}}\sqrt{\frac{3E}{\rho}}\label{Qribbon}
\end{equation}

In another preliminary experiment, tension was applied to the
previously mentioned ribbon of dimensions $0.025\mbox{ mm}$ by
$0.125\mbox{ mm}$ by $0.25\mbox{ m}$. The tension was such that the
resonant frequency of the second violin mode was $\omega = 76\mbox{
Hz}$. The result was $Q_{ribbon} = 18000$, a proof in principal that
the mechanical $Q$ factor of a very thin aluminum ribbon can be
enhanced by applying tension.

The proposed ribbon dimensions for the space DSMG are $0.025\mbox{
mm}$ by $20\mbox{ mm}$ by $1\mbox{ m}$. The proposed ribbon
thickness of $0.025\mbox{ mm}$ is a compromise between the
difficulty of machining a thin ribbon and the difficulty of
supplying enough current to saturate a thick ribbon. The width of
$20\mbox{ mm}$ and length of $1\mbox{ m}$ are the maximum dimensions
that are feasible for a magnetic gradiometer deployed in space. The
DSMG system also requires supporting mechanical structure and
electronics so the dimensions of the entire gradiometer are
approximately $0.04\mbox{ m} \times 0.04\mbox{ m} \times 1.1\mbox{
m}$.

The sensitivity of the DSMG does not directly depend on the $Q$
factor but on the relaxation time $\tau = \frac{2Q}{\omega}$ of the
ribbon. Using Eq. \ref{Qribbon}, the relaxation time is:

\begin{equation}
\tau = Q_{bulk}\frac{l^{2}}{4n\pi
d_{s}}\sqrt{\frac{\rho}{3E}}\label{tauribbon}
\end{equation}

Eq. \ref{tauribbon} shows that $\tau$ is independent of $\omega$.
Nevertheless it is proposed to lower the resonant frequency of the
ribbon from $850\mbox{ Hz}$ to $80\mbox{ Hz}$ in order to keep the
required $Q$ factor to a manageable level. Extrapolating the results
of the preliminary experiment, the increased length of $1\mbox{ m}$
should allow a ribbon $Q$ factor of the order $Q_{ribbon} \approx
300000$ (Q varies as the length of the ribbon squared). The
relaxation time would then be approximately $\tau \approx 1200\mbox{
s}$ which is much greater than its current value of $0.1\mbox{ s}$.
The large increase in the relaxation time of the ribbon should
produce a significant decrease in the thermal noise as per Eq.
\ref{sensitivity}.

\section{Vibration noise}

The existing magnetic gradiometer is designed for airborne
deployment. The high frequency of $850\mbox{ Hz}$ allows a
mechanical isolator to dampen vibration noise from the aircraft by
120dB. In contrast, the proposed DSMG for space deployment would
operate at $80\mbox{ Hz}$, a frequency which would short circuit the
isolator.

The acceleration noise from solar irradiance fluctuations in near
Earth space is on the order of $10^{-10}\mbox{
m}/\mbox{s}^{2}/\sqrt{\mbox{Hz}}$ at $1\mbox{ mHz}$\cite{schumaker}
and less than $10^{-12}\mbox{ m}/\mbox{s}^{2}/\sqrt{\mbox{Hz}}$ at
$80\mbox{ Hz}$\cite{pap}. This compares with seismic noise of $3
\times 10^{-6}\mbox{ m}/\mbox{s}^{2}/\sqrt{\mbox{Hz}}$ on the
ground\cite{coward} and engine noise of $3 \times 10^{-2}\mbox{
m}/\mbox{s}^{2}/\sqrt{\mbox{Hz}}$ on a survey aircraft\cite{mcrae}.

The reduction of vibration noise by 9 orders of magnitude more than
compensates for the greater vibration at lower frequencies. No
vibration isolation is required for the proposed DSMG.

\section{Operation in the natural space environment}

The high current pumped through the ribbon used to detect gradients
generates a significant amount of heat. In a high vacuum the
dominant method of dissipating this heat is thermal radiation. Table
\ref{stress} shows that if the ribbon is allowed to heat up to
$423\mbox{ K}$ then the amount of stress relaxation becomes
unacceptable. It is therefore proposed that the ribbon not be allow
to heat up past $373\mbox{ K}$. For ribbon dimensions as discussed
above this limits the maximum current to $4\mbox{ A}$ (current
density of $8 \times 10^{6}\mbox{ Am}^{-2}$). From Eq.
\ref{sensitivity} it is easy to show that the design sensitivity in
space is $8 \times 10^{-14}\mbox{ T}/\mbox{m}/\sqrt{\mbox{Hz}}$.

\begin{figure}[ht]
\centering
\includegraphics[width = 0.5\textwidth]{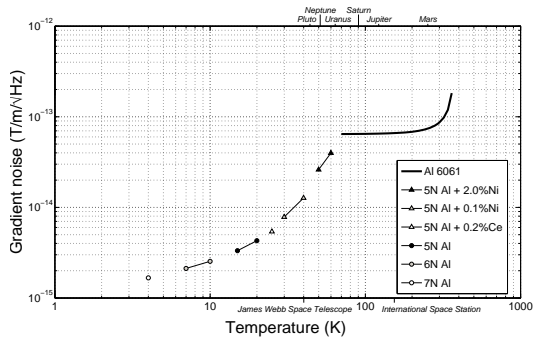}
\caption{String Magnetic Gradiometer design sensitivities as a
function of ambient temperature. There is very little advantage in
cooling the DSMG down to a temperature of $70\mbox{ K}$. Further
decreases in temperature below $70\mbox{ K}$ can however, deliver
large increases in sensitivity. The ambient temperatures of
electronics exposed to sunlight near the outer planets are shown
together with the temperature of some satellites in near earth
orbits which are shielded from the
Sun\cite{nasa}.}\label{sensitivityvstemp}
\end{figure}

The entire ribbon would heat up to $373\mbox{ K}$ and dissipate
approximately $0.8\mbox{ W}$ of power irrespective of the ambient
temperature surrounding the ribbon. The $T^{4}$ power dependence of
thermal radiation means that the DSMG sensitivity depends strongly
on the ribbon temperature yet weakly on the environment temperature.
Fig. \ref{sensitivityvstemp} shows that there is little change in
sensitivity for environment temperatures ranging from $80\mbox{ K}$
to $300\mbox{ K}$.

\section{Thermal radiation in space}

In near Earth space while shielded from the Sun, scientific
instruments radiating their heat into space can reach cryogenic
temperatures between $30\mbox{ K}$ and $50\mbox{ K}$\cite{stsi}. The
advanced sun shields that have been manufactured for the James Webb
Space Telescope can reduce $1370\mbox{ Wm$^{-2}$}$ of sunlight
impacting on the front of the shield down to a mere
$105\mbox{$\mu$Wm$^{-2}$}$ behind the shield where the ambient
temperature is approximately $7\mbox{ K}$\cite{stsi}\cite{amato}. A
proposed ribbon made from low resistivity aluminium alloys
(described in the next section) would dissipate a tiny $30\mbox{
mW}$ of power despite a high current of $10\mbox{ A}$. High current
audio amplifiers are available commercially with output impedances
as low as $0.005\mbox{ $\Omega$}$. Heat dissipated from power
supplies and support electronics could be screened from the ribbon
by highly reflective mirrors or by displacing all heat producing
DSMG modules a meter from the ribbon.

These figures show that passive cooling of the DSMG system in space
is feasible. A radiator area of $0.2\mbox{ m$^{2}$}$ would cool the
ribbon down to $40\mbox{ K}$, whilst a radiator area of $1\mbox{
m$^{2}$}$ would cool the system down to $27\mbox{ K}$.

\section{25 Kelvin to 70 Kelvin operation}

\begin{figure}[ht]
\centering
\includegraphics[width = 0.5\textwidth]{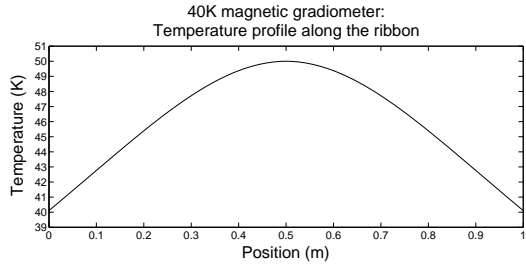}
\caption{$0.03\mbox{ W}$ of power is dissipated by thermal
conduction along the $1\mbox{ m}$ length of the ribbon. The middle
of the ribbon heats up to $50\mbox{ K}$ while the ends are in
thermal contact with a heat sink at $40\mbox{ K}$.}\label{40k}
\end{figure}

At $40\mbox{ K}$ the electrical resistivity of high purity 99.999\%
5N aluminium is very low $\rho \approx 1.80 \times 10^{-10}\mbox{
}\Omega\mbox{m}$ and the thermal conductivity very high $\kappa
\approx 2000\mbox{ W}/\mbox{m}/\mbox{K}$. The thermal conductivity
values are from an empirical fit by Woodcraft\cite{woodcraft} and
the temperature dependence of resistivity are from Hashimoto et
al\cite{hashimoto2}. These properties make thermal conduction along
the axis of the ribbon the dominant form of heat dissipation below
$70\mbox{ K}$. Fig. \ref{40k} shows the temperature profile along
the ribbon for an ambient temperature of $40\mbox{ K}$.

At temperatures below 10\% of the melting point of aluminium, work
hardening reduces creep to zero\cite{frost} so annealed alloys with
low resistivity can be used. Switching the ribbon material from
Aluminium 6061 to pure aluminium may be unfeasible since the yield
strength of 5N aluminium is only $5\mbox{ MPa}$\cite{reed}. Instead
it is proposed to use special highly conductive high strength alloys
such as those developed for the Atlas Project\cite{wada}\cite{wada2}
as shown in table \ref{materials}. One alloy of 5N aluminium with
0.1\% nickel has a greatly increased yield strength of $80\mbox{
MPa}$ at a cost of only moderately higher electrical resistivity
($\rho \approx 2.2 \times 10^{-10}\mbox{ }\Omega\mbox{m}$ at
$40\mbox{ K}$).

The high thermal and electrical conductivities mean that the current
could be increased to $10\mbox{ A}$ and the ribbon width reduced to
$18\mbox{ mm}$. The current density could be as high as $2 \times
10^{7}\mbox{ Am}^{-2}$. Eq. \ref{sensitivity} shows that high
currents, low ribbon mass and low temperature increase the signal to
noise ratio. The design sensitivity is approximately $10^{-14}\mbox{
T}/\mbox{m}/\sqrt{\mbox{Hz}}$ at $40\mbox{ K}$.

\section{4 Kelvin to 25 Kelvin operation}

Further increases in sensitivity are possible if the temperature
could be lowered below $25\mbox{ K}$ by utilizing very large thermal
radiators in conjunction with sun shields or by using liquid helium.
As shown in table \ref{materials}, first 5N, then 6N\cite{reed} and
eventually 7N single-crystal aluminium\cite{hashimoto} are necessary
to exploit these low temperatures.

For pure metal films, the resistance drops with decreasing
temperatures until the mean free path $l$ of conducting electrons
becomes larger than the film thickness $t$. In the limit $l
>> t$ the ribbon resistance will increase by a factor
of\cite{sondheimer}:

\begin{equation}\frac{R_{ribbon}}{R_{bulk}} =
\frac{4l}{3t\log(l/t)}
\end{equation}

Using the empirical relationship $l\rho = 8.2 \times 10^{-16}\mbox{
}\Omega \mbox{m}^{2}$\cite{sambles}, the mean free path of 7N
aluminium with a resistivity of $\rho = 1.5 \times 10^{-13}\mbox{
}\Omega\mbox{m}$ is $l = 5.4\mbox{ mm}$. Magnetoresistance reduces
the mean free path to $4.7\mbox{ mm}$\cite{egan}. This value of the
mean free path is $187$ times larger than the thickness of the
ribbon. The resistance of the ribbon would then increase by a factor
of $47$ due to this size effect. Size effects continue to be
significant for temperatures up to $25\mbox{ K}$.

The low electrical resistivity at liquid helium temperatures can not
be fully realised due to size effects. Other difficulties with
exploiting a low temperature environment include magnetoresistance,
self inductance and the low strength of pure alloys. Fig.
\ref{sensitivityvstemp} shows that the magnetic gradient sensitivity
tapers off below 10K. The design sensitivity at $4\mbox{ K}$ is
approximately $2 \times 10^{-15}\mbox{ T}/\mbox{m}/\sqrt{\mbox{Hz}}$
using a current of $10\mbox{ A}$ and a current density of $2 \times
10^{8}\mbox{ Am}^{-2}$. The small improvement in sensitivity does
not justify the difficulty of achieving liquid helium temperatures.
DSMGs are only feasible for temperatures between $25\mbox{ K}$ and
$300\mbox{ K}$.

\section{Low frequency noise background}

"The phenomenon of $1/f$ noise, with spectral density scaling
inversely with frequency is common to virtually all
devices"\cite{koch}. The typical frequencies of interest in for
global magnetic surveys range from $200\mbox{ }\mu\mbox{Hz}$ to
$0.17\mbox{ Hz}$\cite{hastings}. Deep space missions such as the
voyager measure the slowly varying interplanetary magnetic field
with frequencies ranging from $50\mbox{ nHz}$ up to $1\mbox{
Hz}$\cite{burlaga}. $1/f$ flicker noise and random walk noise are
expected to be significant at these frequencies. The low frequency
noise of the DSMG is compared with some existing devices in this
section.

Fluxgate magnetometers have been used in space for more than 30
years. The noise power spectral density of a high performance
fluxgate magnetometer is typical of shot noise devices and
characterised by a $1/f$ spectrum with a typical value of $3 \times
10^{-12}T/\sqrt{\mbox{ Hz}}$ at $1\mbox{ Hz}$ for space
research-grade instruments\cite{primdahl}. The source of the noise
is attributed to Barkhausen-like mechanisms that affect the motion
of domains in ferromagnetic material in the sensor
cores\cite{acuna}. One fluxgate magnetic gradiometer built for use
in space has a sensitivity of $3 \times 10^{-11}\mbox{
T}/\mbox{m}/\sqrt{\mbox{Hz}}$ at $1\mbox{ Hz}$\cite{merayo}.

Low $T_{c}$ SQUIDs were proposed as a highly sensitive magnetic
gradiometer for space-borne magnetic investigations more than 20
years ago\cite{hastings}. SQUIDs have the best sensitivity on offer
for terrestrial operations. One low $T_{c}$ SQUID gradiometer has a
sensitivity of $6 \times 10^{-14}\mbox{
T}/\mbox{m}/\sqrt{\mbox{Hz}}$\cite{stolz} in the laboratory with a
$1/f$ noise corner at $0.3\mbox{ Hz}$. $1/f$ flicker noise is more
severe in high $T_{c}$ gradiometers which have a typical $1/f$ noise
corner of $10$Hz\cite{zhang}\cite{tilbrook}. The two major sources
of $1/f$ noise in dc SQUIDs are fluctuations of the critical current
in the Josephson junctions and motion of flux lines (vortices)
trapped in the body of the SQUID\cite{fagaly}. The frequency ranges
of interest in space exploration are entirely within the $1/f$ noise
region of SQUIDs\cite{hastings}.

Previously, all existing gradiometers measured the gradient from the
differential output of two sensing elements. Uniform fields
contaminate the results since the common mode rejection ratio of
gradiometers is finite. Typically the common mode rejection ratio is
on the order of $10^{4}$\cite{stolz}\cite{zhang}\cite{bick} although
rotating gradiometers can do better\cite{tilbrook}. The common mode
rejection ratio tends to decrease with time\cite{bick} which
introduces additional noise at very low frequencies.

The noise performance of magnetometers is strongly affected by
mechanical and thermal stresses, which vary over time and with
exposure to extreme environments. Even state of the art
magnetometers experience drift on the order of $10^{-10}\mbox{
T}/\mbox{year}$\cite{acuna}. These mechanical and thermal effects
introduce a random walk noise with power scaling $1/f^{2}$.

Because the DSMG is a single element device, any drifts associated
with non-stationary unbalance of two different sensors (as in the
case of multisensor based gradiometers) are absent. Also, it is a
modulation-demodulation device such that most of the $1/f$ noise in
the DC region is suppressed compared to static gradiometers.

\begin{figure}[ht]
\centering
\includegraphics[width = 0.5\textwidth]{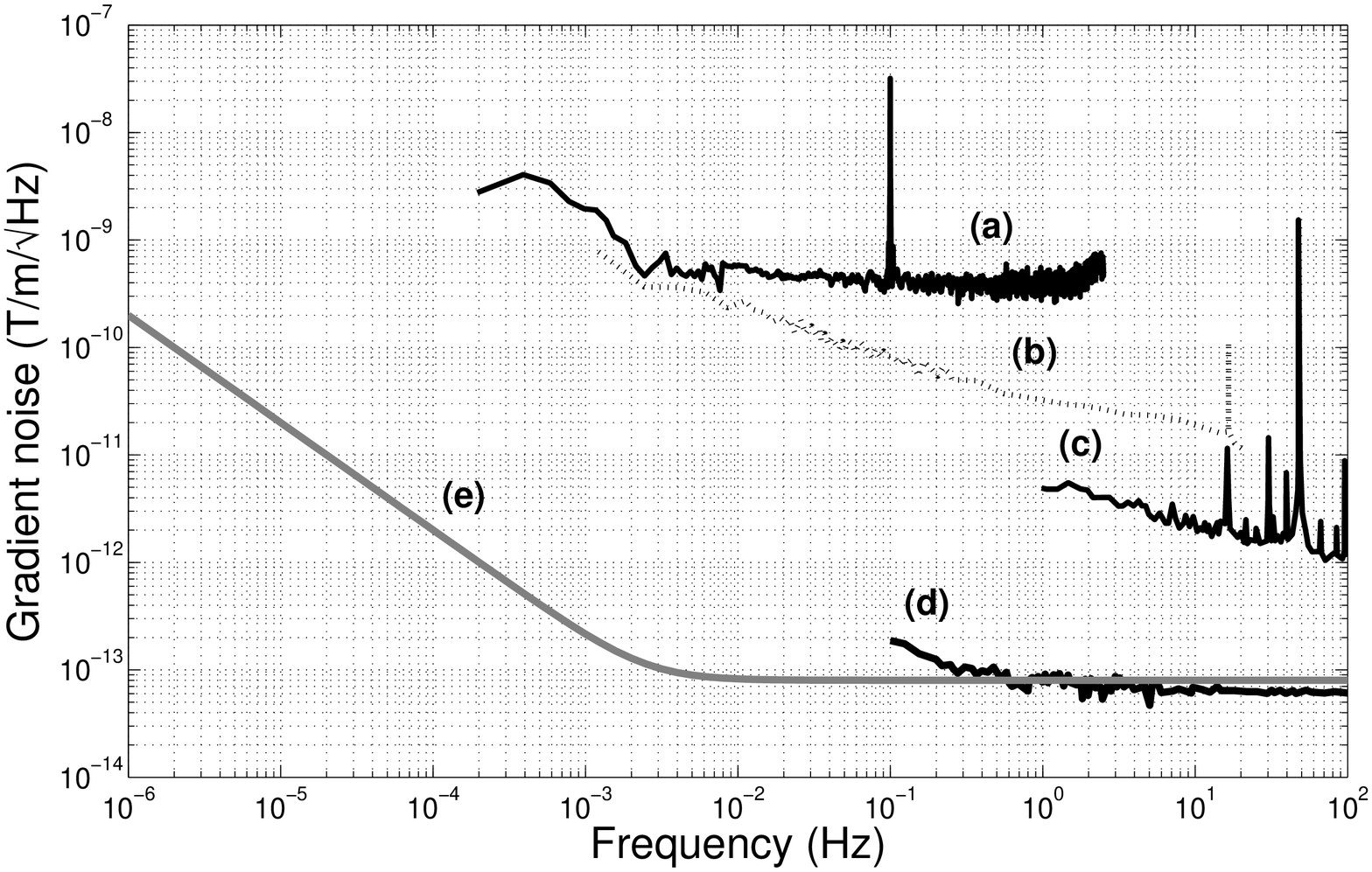}
\caption{The magnetic gradient system noise of (a) an existing DSMG
operating in a moderate vacuum of $1\mbox{ kPa}$ in unshielded
environment, (b) a fluxgate gradiometer developed for magnetic
surveying from a satellite\cite{merayo}, (c) a high temperature
SQUID gradiometer developed for magnetocardiography\cite{zhang}, (d)
a low temperature SQUID gradiometer developed for airborne mineral
exploration\cite{stolz} and (e) the proposed space DSMG in high
vacuum.}\label{lowfrequencynoise}
\end{figure}

The low frequency noise of the DSMG is characterised by a $1/f^{2}$
spectrum of random walk noise. In the closed loop
operation\cite{veryaskin2}, the ground based DSMG has a $1/f^{2}$
noise corner of approximately $2.5\mbox{ mHz}$ (see Fig.
\ref{lowfrequencynoise}). With the feedback loop turned off, the
frequency of the noise corner depends on the magnitude of the
external gradient; noise corners as high as $0.2\mbox{ Hz}$ have
been measured when the DSMG is exposed to gradients higher than
$10000\mbox{ nT/m}$.

\section{Low frequency noise improvements}

The white noise floor of the space DSMG is lower than the ground
based DSMG by several orders of magnitude. If the level of random
walk noise remains the same then the frequency of the $1/f^{2}$
noise corner would increase. The origins of the low frequency noise
in the DSMG are not known at present although several models have
been proposed. However, there are reasons for believing that the
level of low frequency noise can be reduced step in step with the
white noise.

Any mismatch between the drive frequency of the AC current and the
resonant ribbon frequency creates low frequency noise. The space
DSMG's low level of white noise would allow more accurate tracking
of the ribbon resonant frequency. In addition, the amplitude of
ribbon vibrations during closed loop operation could be reduced step
in step with the white noise. By using the feedback of closed loop
operation to reduce the signal\cite{veryaskin2}, any amplitude
dependant low frequency noise will be reduced accordingly.

The very high mechanical Q factor of the space DSMG could be
utilised for off resonance operation. The very large signal at
resonance could be used exclusively to identify the exact resonant
frequency of the ribbon. The level of signal at resonance could be
ignored. The difference between the AC current drive frequency and
the resonant frequency would be known exactly. Amplitude
fluctuations from frequency drift could be fully compensated while
any small variance in the mechanical Q would produce only negligible
noise during off resonance operation.

For the reasons outlined above, the noise spectrum of the proposed
DSMG for space applications is expected to be flat for frequencies
down to $2.5\mbox{ mHz}$. Fig. \ref{lowfrequencynoise} compares the
expected performance of the DSMG with other technologies and shows
predominance of $1/f$ noise and $1/f^{2}$ noise at low frequencies.

\section{Readout}

In addition to the thermal noise in the ribbon, there is also
measurement noise in the apparatus used to measure the position of
the ribbon. In the limit $Q \rightarrow \infty$, the rms
displacement $x$ of the ribbon produced by a magnetic gradient is:

\begin{equation}
x = \frac{dB_{y}}{dz}\frac{l i}{n\pi^{3}\sqrt{32}\eta f_{0}f}
\end{equation}

where $n = 2$ is the mode number, $l$ is the length of the ribbon,
$\eta$ is the mass per unit length, $i$ is the peak current, $f_{0}$
is the resonant frequency of the 2nd violin mode of the ribbon and
$f$ is the frequency of the magnetic gradient signal. For ground
based operations, the smallest detectable signal of $4 \times
10^{-10}\mbox{ T}/\mbox{m}/\sqrt{\mbox{Hz}}$ produces a displacement
of $10^{-11}\mbox{ m}/\sqrt{\mbox{Hz}}$ over a bandwidth of $1\mbox{
Hz}$.

The present method of measuring ribbon deflections is pumping the
ribbon with a small radio frequency current of $i = 0.2\mbox{ A}$.
The radio frequency current generates a radio frequency flux around
the ribbon. Two pickup coils connected in differential mode measure
the modulation in flux as the ribbon moves. Each pickup coil is
capable of measuring ribbon deflections of size $6 \times
10^{-13}\mbox{ m}/\sqrt{\mbox{Hz}}$ which is sufficient to detect
the minimum signal.

If the sensitivity improvements suggested in this paper are
implemented then the minimum signal amplitude will be only $9 \times
10^{-15}\mbox{ m}/\sqrt{\mbox{Hz}}$ for a $1\mbox{ Hz}$ signal. In
order to improve the readout sensitivity enough to detect a
displacement this small, the pickup coils could be replaced with a
microwave cavity readout developed for gravitational wave
antennas\cite{ivanov}. Other solutions include a low noise
SQUID\cite{carelli} or optical readout using a Fabry Perot
cavity\cite{conti}.

\section{Conclusion}

The proposed space DSMG has a sensitivity of $8 \times
10^{-14}\mbox{ T}/\mbox{m}/\sqrt{\mbox{Hz}}$ using only the natural
space environment. Even higher sensitivity is possible with passive
cooling of the DSMG in space. The size of the DSMG and the power
consumption requirements are comparable with existing magnetometers
used in space missions. It is also possible to deploy a full tensor
gradiometer by combining several single axis DSMGs.

\ack The authors would like to thank Mr. Howard Golden of Gravitec
Instruments for many useful discussions and suggestions. Work on the
DSMG project is funded in part by a linkage grant from the
Australian Research Council.



\begin{table}[ht]
\centering
\includegraphics[width = 0.32\textwidth]{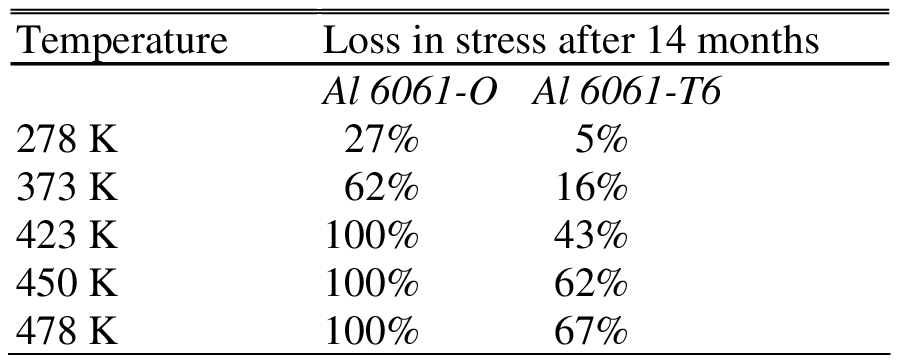}
\caption{\rm{The results in this table are taken from
Kaufman\cite{kaufman}. Stress relaxation becomes increasingly
problematic at higher temperatures. Tempering the aluminium (Al
6061-T6) can reduce stress relaxation compared with fully annealed
aluminium (Al 6061-O).}} \label{stress}
\end{table}

\begin{table}[ht]
\centering
\includegraphics[width = 0.5\textwidth]{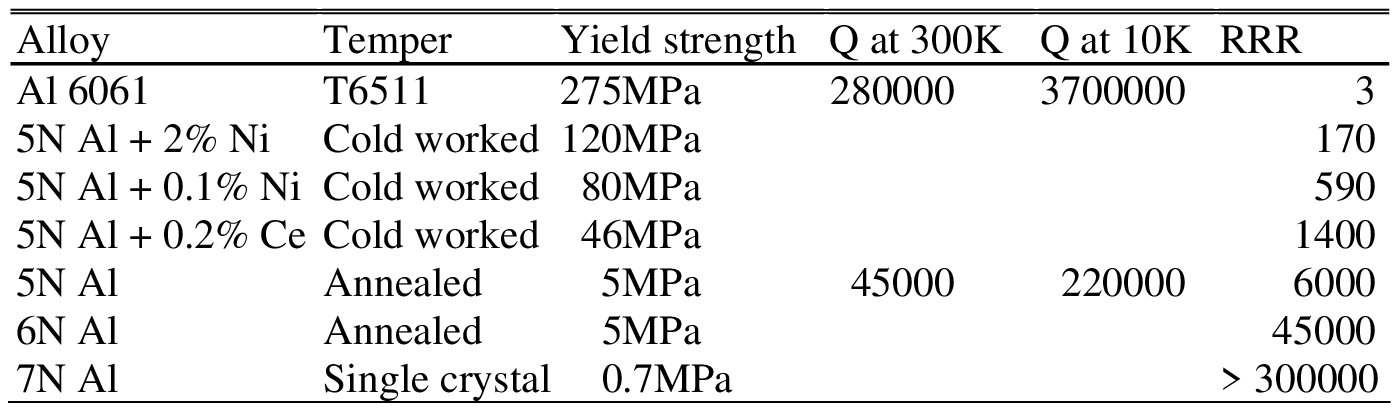}
\caption{\rm{At progressively lower temperatures, aluminium
alloys\cite{kaufman}\cite{wada}\cite{reed}\cite{hashimoto} with
lower and lower values of electrical resistance are proposed. The
low resistance comes at the expense of a reduced yield strength. RRR
(residual resistivity ratio) is the ratio of the electrical
resistivity at room temperature to the residual resistivity from
impurities at liquid helium temperatures.}} \label{materials}
\end{table}

\end{document}